\documentclass [{onecollarge,final}] {svjour2}

\usepackage{graphicx}
\usepackage{dcolumn}
\usepackage{bm}
\usepackage{hyperref}
\usepackage{amssymb}
\usepackage{hyperref}
\usepackage{braket}
\RequirePackage[normalem]{ulem} %DIF PREAMBLE
\RequirePackage{color}\definecolor{RED}{rgb}{1,0,0}\definecolor{BLUE}{rgb}{0,0,1} %DIF PREAMBLE
\usepackage{mathtools} \usepackage{amsmath}

\usepackage{graphicx}% Include figure files
\usepackage{dcolumn}% Align table columns on decimal point
\usepackage{bm}% bold math
\usepackage{hyperref}% add hypertext capabilities
\usepackage{amssymb}%\usepackage[mathlines]{lineno}% Enable numbering of text and display math
\usepackage{hyperref}% add hypertext capabilities
\RequirePackage[normalem]{ulem} %DIF PREAMBLE
\RequirePackage{color}\definecolor{RED}{rgb}{1,0,0}\definecolor{BLUE}{rgb}{0,0,1} %DIF PREAMBLE
\usepackage{mathtools} \usepackage{amsmath}

\usepackage{graphicx}
\usepackage{amssymb}
\usepackage{amsmath}
\usepackage{bm}
\usepackage{enumitem}

\usepackage{graphicx}
\usepackage{amssymb}
\usepackage{amsmath}
\usepackage{bm}
\usepackage{enumitem}

\usepackage{graphicx}
\usepackage{dcolumn}
\usepackage{bm}
\usepackage{hyperref}
\usepackage{amssymb}
\usepackage{hyperref}
\usepackage{graphicx}
\usepackage{hyperref}
\begin{document}

\title{On the Exponential   Decay of   Strongly Interacting Cold Atoms from a Double-Well Potential}
\author{Przemys\l aw Ko\'{s}cik}
\institute{University of Applied Sciences, Department of Computer Sciences,
 ul. Mickiewicza 8, PL-33100 Tarn\'{o}w, Poland}

\maketitle

\begin{abstract}
In this article, we study an exponential decay for the gas of bosons with strong repulsive delta interactions from a double-well potential. We consider an exactly solvable model comprising an infinite wall and two Dirac delta barriers. 
 We explore its features both within the exact method and with the resonance expansion approach. The study reveals  the effect of  the   splitting barrier   on 
the decay rate in dependence on the number of particles.
Among other things, we find that the effect of the splitting barrier on the decay rate  is most pronounced in systems with odd particle numbers. During   exponential decay, the spatial  correlations in an internal region     are  well captured by the  "radiating  state".
\end{abstract}

\section{ Introduction}

Over the past few years,  there has been a growing interest in understanding the decay properties of unstable quantum states     \cite{alpha,alpha1,Winter0,garcia,wy,Fran,for,romo}. In particular,  recent progress in  fabricating  systems of interacting particles has inspired the theoretical community to  study  the decay properties of unstable  many-particle states \cite{two,sow,two1,exp,TG1,kos}. A simple model to study the decay process of many-particle states is the system of bosons with infinitely strong delta-contact interactions, i.e., the so-called Tonks-Girardeau (TG) gas \cite{Girardeau}.  Considerable effort has been made already to understand the decay properties of such systems. Among other works, the relevant for exponential and long-time decays were presented in \cite{exp} and  \cite{TG1}, respectively.   A recent paper \cite{kos} went even further and explained the decay mechanism of TG gases at intermediate stages of the time evolution (between exponential and long-time regimes).

\begin{figure}
\includegraphics[width=90mm]{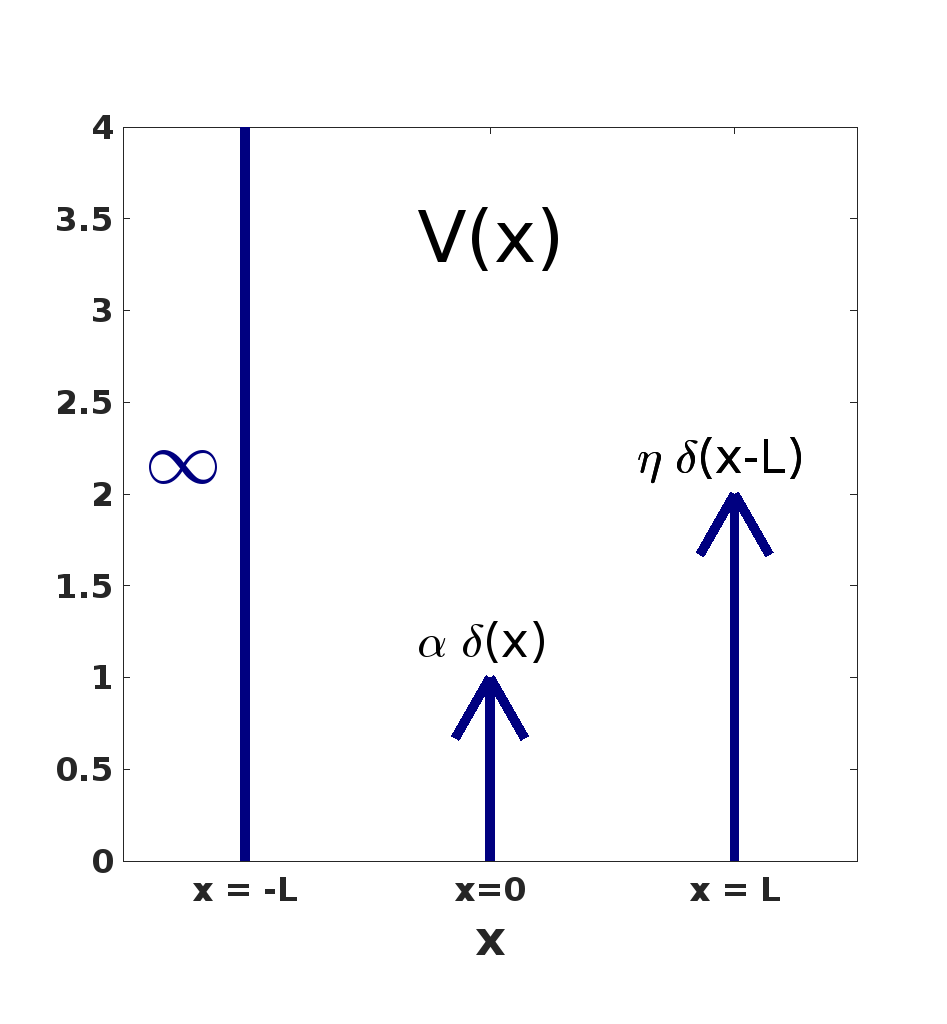}
\caption{Ilustrative diagram of double-well structure in Eq. (\ref{pot})\label{Fig1}}
  \end{figure}

    In the present paper, we provide a deeper insight into the exponential decay of TG gases from the double-well trap. As a model, we consider the potential in the form \cite{kos}

 \begin{equation}\label{pot} V(x)=\begin{cases} \infty, & \text{if $x\leq-L$}\\\alpha\delta(x)+\eta\delta(x-L) & \text{otherwise}. \end{cases} \end{equation} 
Note that the model is a modification 
of the celebrated \textit{Winter } model \cite{Winter0}, to which   a   Delta  barrier at $x=0$ was added, see Fig.\ref{Fig1}. The remainder of this article is structured as follows.  Section \ref{results} discusses the  theoretical tools for studying the time evolution of the decaying  TG gas   and focuses on the results.   
Section \ref{conclusion} presents some concluding remarks.

\section{Results}\label{results}
 The scenario we consider is typical of controllable studies of the 
tunnelling phenomena in modern experiments. In the case studied, the system is initially  prepared $(t<
 0)$ in the ground state of the TG gas in a hard-wall split trap ($\eta=\infty$). At $t=0$, the strength of the right barrier is changed to a finite value of $\eta$. As a result, the initial state is no longer stationary and  begins to evaluate in time. According to  Bose-Fermi mapping \cite{Girardeau}, the time-dependent TG wave function is given by
\begin{eqnarray}\label{tgfun}\Psi(x_{1},x_{2},...,x_{N},t)=\nonumber\\=\mathrm{\Pi}_{k<l}\mathrm{sgn}(x_{k}-x_{l}){1\over \sqrt{N!}}\mathrm{det}_{i,j}^{N}[\phi_{i}(x_{j},t)], \end{eqnarray} where the one-particle state  $\phi_{k}(x,t)$ is governed by the  Schr{\"o}dinger equation,

 \begin{equation}\label{sh1}
 I\hbar {   \partial \phi_{k}(x,t)\over \partial t}= [-\frac{\hbar ^2}{2m}{\partial^2\over \partial x^2}+ V(x)]\phi_{k}(x,t),
\end{equation}
with the initial condition as  the bound-state eigenfunction of the  hard-wall split trap, $\varphi_{k}(x)$, that is, $\phi_{k}(x,t)|_{t=0}=\varphi_{k}(x)$.
From here we set $L=\hbar= m=1$ so that the  spatial coordinates,  time  coordinates, and  energies are measured in units of $L$ and $mL^2/\hbar$,   in and $\hbar^2/(mL^2)$, respectively.
The system under  consideration has a nice feature where   both  eigenfunctions  of the  hard-wall  split trap, $\varphi_{k}(x)$  and continuum    wave functions  ($\eta<\infty$)  $\psi_{p}(x)$ (normalised to a delta Dirac distribution) can be obtained in  closed analytical forms.    For further detail,  we refer  readers to the papers \cite{kos,anal}, in which   the relevant formulas are reported.
 Thanks to those,  
the solutions to Eq. (\ref{sh1}) can be condensed    in   a Fourier series as follows:
\begin{equation}\label{time}\phi_{k}(x,t)=\int_{0}^{\infty} c_{k}(p)\psi_p(x)e^{-{I t p^2\over 2} }dp,\end{equation}
where  $c_{k}(p)$, 
\begin{equation}\label{mom} c_{k}(p)=\int_{-1}^{1}{ \varphi}_{k}(x) \psi_p(x)dx,\end{equation} is given in closed analytical form \cite{kos}. Nonetheless,  numerical computations are required to evaluate the integrals in
 Eq. (\ref{time}). 
 We conduct our analysis  in terms of the non-escape probability,
\begin{equation}\label{nonex} 
P^{(N)}(t)=\int_{\Delta^{N}}|\Psi(x_{1},x_{2},...,x_{N},t)|^2dx_{1}...dx_{N},
\end{equation}  
$\Delta^{N}=[-1,1]^N$ (internal region), which informs us of the probability that $N$ bosons remain in   the internal region  at time $t$.
 For the TG wavefunction in  Eq. (\ref{tgfun}), the  non-escape probability can be  reduced to the  matrix form \cite{exp,TG1} 
$P^{(N)}(t)=\mathrm{det}_{k,l}^{N} [P_{kl}(t)]$
with the entries
$P_{kl}(t)=\int_{-1}^{1} \phi^{*}_{k}(x,t)\phi_{l}(x,t)dx$. The diagonal elements $P_{kk}(t)$ are nothing but  the non-escape probabilities of the  one-particle states. We denote 
$P_{k}(t)=P_{kk}(t)$.
  Within the resonance expansion method \cite{garcia},  the one-particle state that experiences an exponential decay follows   the approximation:
  $\phi_{k}(x,t)\approx M_{k}(x)e^{-\Gamma_{k}t/2-I  \varepsilon_{k}t}$ with $\Gamma_{k}=-\mathrm{Im}\{p_{k}^2\}$,
   $\varepsilon_{k}=(\mathrm{Im}\{p_{k}\}^2-\mathrm{Re}\{p_{k}\}^2)/2$, and $M_{k}(x)=2\pi I res_{p_{k}}\{c_{k}(p)\psi_{p}(x)\}$, where $p_{k}$ are  the roots of the denominator in the integrand in Eq.(\ref{time})  on  the fourth quadrant of the complex $p$-plane (proper poles)  and
    $res_{p_{k}}\{f\}$ stands for the residue of $f$ at a pole $p_{k}$. It should be mentioned that numerical calculations are needed to obtain the values of $p_{k}$. The corresponding  non-escape probability  is 
 $P_{k}(t)\approx m_{k}e^{-\Gamma_{k}t}$, where $m_{k}=\int_{-1}^{1}|M_{k}(x)|^2dx$.
If $m_{k}\approx 1$($P_{k}(0)\approx1$), then the exponential decay starts at about $t=0$ and 
  the state $\ket{\phi_{k}(t)}$  can well be  approximated in the region $\Delta$ by the so-called "radiating state":    $\ket{\phi_{k}(t)}\approx e^{-\Gamma_{k}t/2-I \varepsilon_{k}t}\ket {\phi_{  k}(0)}$.  When taking this approach, the time-dependent TG wavefunction  in the region $\Delta^{N}$ takes the form:
  \begin{eqnarray}\label{tfun}\Psi(x_{1},x_{2},...,x_{N},t)\approx e^{-\Gamma^{(N)}t/2-I\varepsilon^{(N)} t }\mathrm{\Pi}_{k<l}\mathrm{sgn}(x_{k}-x_{l}){1\over \sqrt{N!}}\mathrm{det}_{i,j}^{N}[\phi_{i}(x_{j},0)], \end{eqnarray}
 with $\Gamma^{(N)}=\sum_{k=1}^N \Gamma_{k}$,
$\varepsilon^{(N)}=\sum_{k=1}^{N}\varepsilon_{k}$. Consequently, its validity is expected to hold when  $m_{k}\approx 1$ for $k=1,...,N$ (
 $M^{(N)}=\Pi_{k=1}^{N}m_{k}\approx 1$).
 The corresponding $N$-particle non-escape probability   is:   \begin{equation}\label{tg1}P^{(N)}(t)\approx e^{-\Gamma^{(N)}t}. \end{equation} 
Now, we focus on examining when the above approximations are satisfied and what follows from their applicability.
\begin{figure}
\includegraphics[width=83mm]{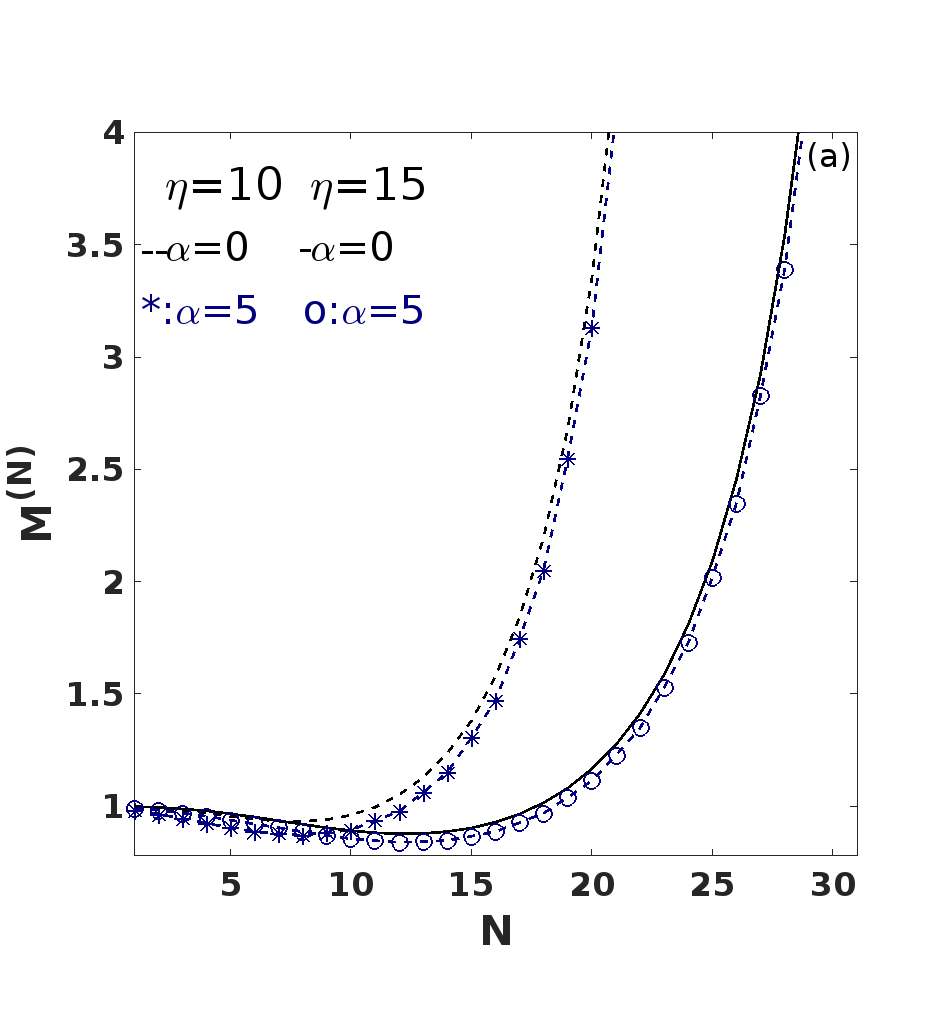}
\includegraphics[width=83mm]{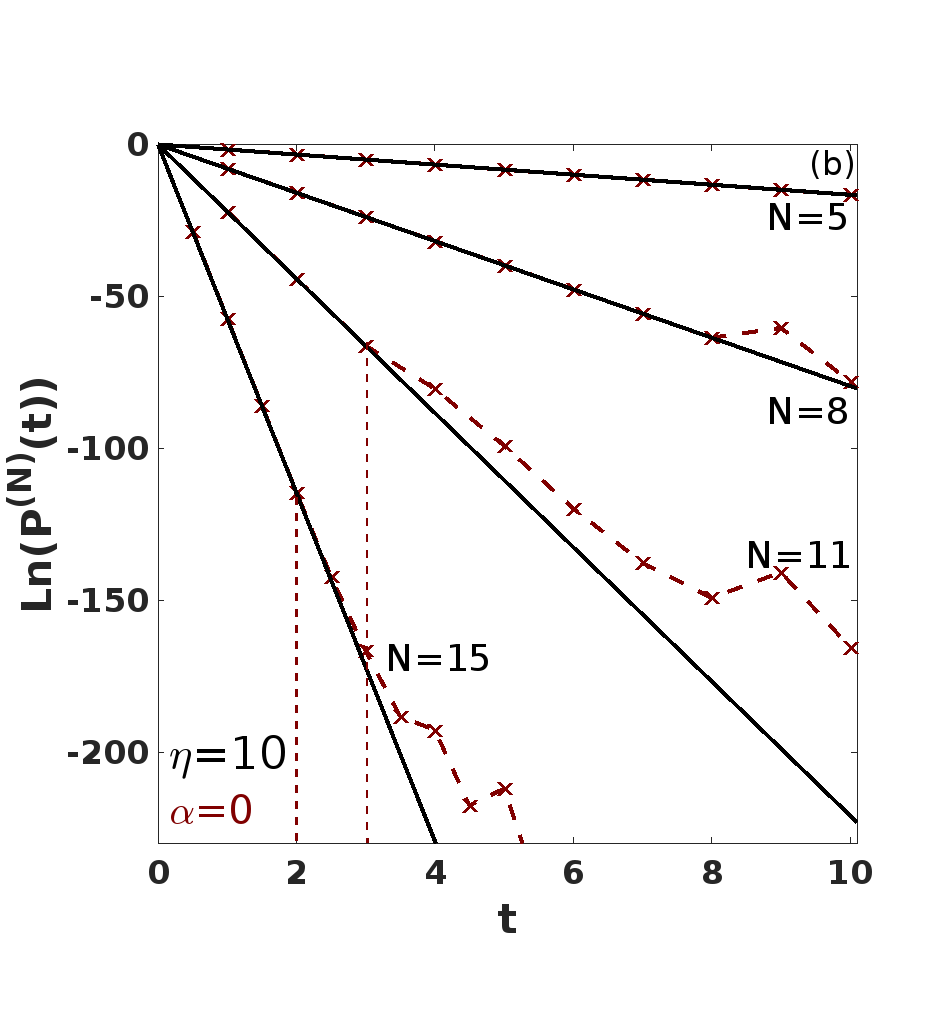}
\includegraphics[width=83mm]{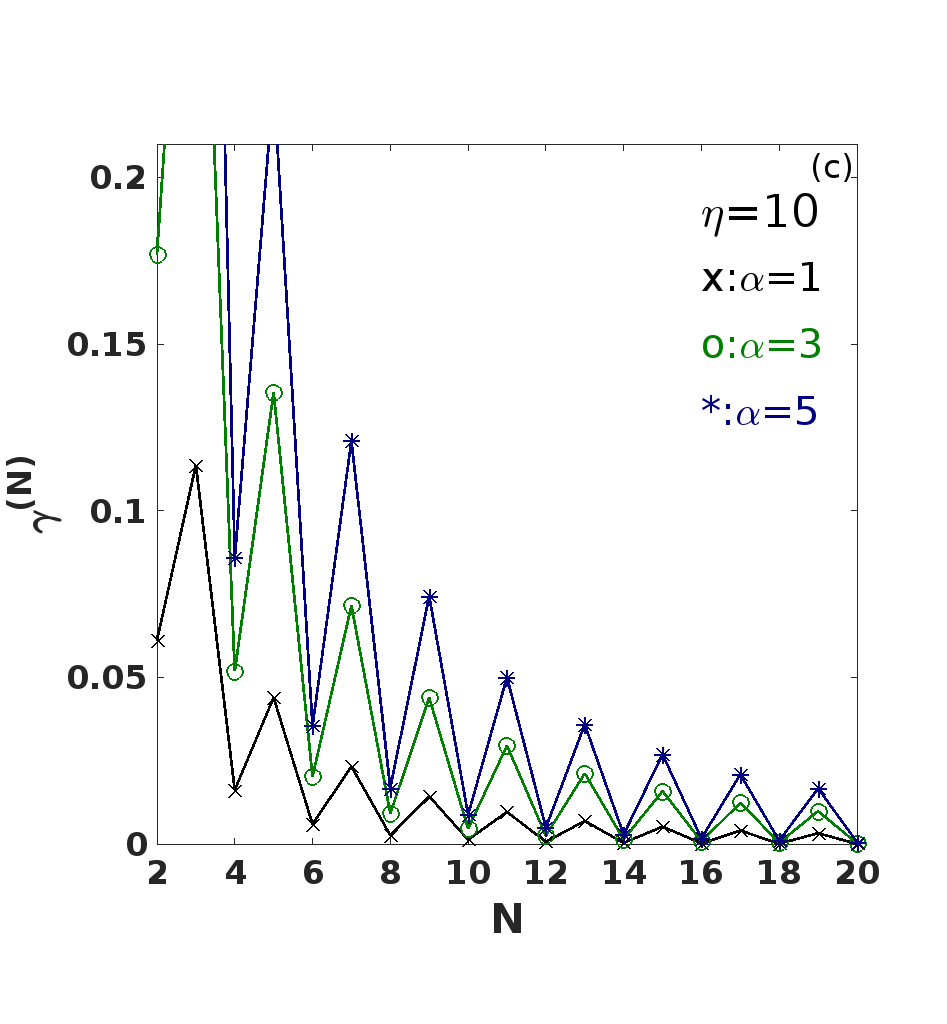}
\includegraphics[width=83mm]{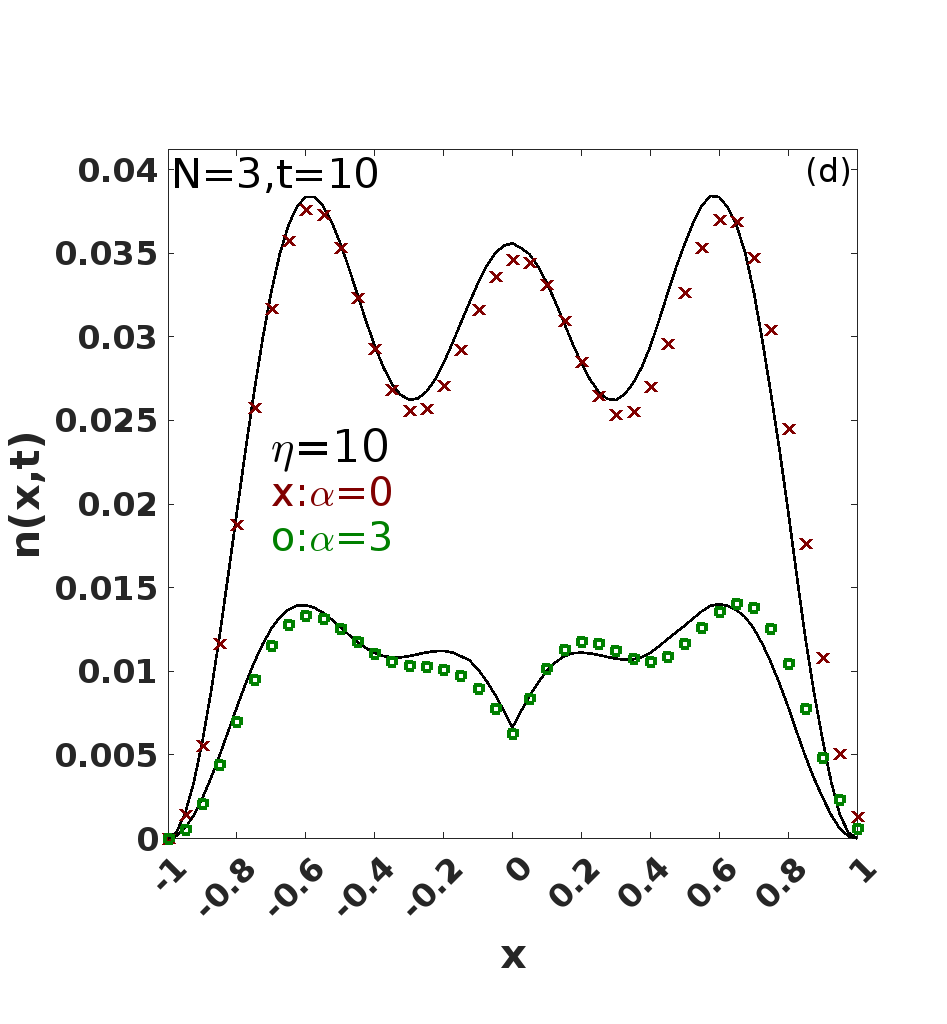}
\caption{Results for  different quantities as discussed in the text obtained for some    transparent control parameter values.  (a)  Results for $M^{(N)}$ as a function of $N$. (b)    $N$- particle non-escape probability, where the continuous lines represent 
the results of  Eq.(\ref{tg1}) and the markers  the results of \textit{exact} calculations. For clarity, the dashed vertical lines  mark the points where the deviations  from the exponential decay begin in the cases of $N=11$ and $N=15$. (c) The behaviour of the relative change $\gamma^{(N)}$. (d)  Results for $n(x,t)$,  obtained for $N=3$ particles at $t=10$ from the numerically \textit{exact} wave function (markers) and its approximation in Eq. (\ref{tfun})(continuous lines)\label{Fig2} }
\end{figure}
 Our results are summarised in Fig. \ref{Fig2}.
  Fig. \ref{Fig2} (a) shows the behaviour of     $M^{(N)}$ as a function of $N$. We can observe a local minimum   with a value slightly smaller than one. As a result, there is a  value where $N=N_{c}$(greater than where the minimum occurs) such that $M^{(N_{c})}\approx 1$. 
  Starting from $N=N_{c}$  the deviation of  $M^{(N)} $from $1$ rapidly  increases  as $N$ increases.
This suggests that the point $N=N_{c}$
can be viewed as a transition point to the regime in which the non-escape probability begins to diverge significantly from its approximation  in  Eq. (\ref{tg1}).
  To clarify this, Fig. \ref{Fig2} (b)  offers a comparison of the results obtained from Eq. (\ref{tg1}) with the results of  the \textit{exact} numerical calculations, where to support the presentation only  case $\eta=10,\alpha=0$ is shown. 
As the results indicate, the  period in which the decay is consistent with  the "radiating state"     shrinks with increasing $N$. When $N$ exceeds the 
critical value $N_{c}=10$ (see Fig. \ref{Fig2} (a)), the  decay of the $N$-particle state switches to the non-exponential regime  at a very small $t$ value. 
When the splitting barrier is present
an effect of the parity of a number of particles appears.
This is  demonstrated in  Fig. \ref{Fig2} (c) which displays the behaviour of   a  relative change defined as  $\gamma^{(N)}=
(\Gamma^{(N)}-\Gamma^{(N)}_{\alpha=0}):\Gamma^{(N)}_{\alpha=0}$, where  $\Gamma^{(N)}_{\alpha=0}$ represents the 
decay rate for a system without the splitting barrier.
We conclude that the change in the decay process caused by the addition of the splitting barrier is most pronounced for systems with odd numbers of particles and   in the  small   $N$ regime, i. e.  where $\gamma^{(N)}$ exhibits its most rapid variation.    
 When there is instead an even 
number of particles, the decay rate becomes almost insensitive to changes in  $\alpha$.
 It is worth mentioning that    the coherence of the initial state  depends on $\alpha$ in the opposite way  \cite{anal}. 
 That 
is to say, it strongly depends on $\alpha$ only when $N$ is even.
To determine whether  Eq. (\ref{tfun}) can capture  the correlation in the internal region, we  tested  its ability to reproduce a function   $n(x,t)$ defined as $P^{(N)}(t)=\int_{-1}^{1}n(x,t)dx$,  
\begin{equation}\label{dex} 
n(x,t)=\int_{\Delta^{N-1}}|\Psi(x,x_{2},...,x_{N},t)|^2dx_{2}...dx_{N}.
\end{equation}  
 If  $n(x,t)$ is divided   by the corresponding value of  $P^{(N)}(t)$, then the resulting  quantity  can be interpreted as the  probability density of finding a particle, provided that     all the particles  remain in the region $\Delta$. 
 The correctness of the "radiating state" in reproducing the spatial correlation   is confirmed in Fig. \ref{Fig2} (d), where the results for $n(x,t)$ obtained using it and the  \textit{exact} wave function  are compared.
 The results imply  that during exponential decay,  spatial correlations in the internal region are indeed consistent with  those in the initial state.   More  details regarding  the correlation in the initial state (i.e. in the TG ground state in the hard-wall split trap) can be found in \cite{anal}.

\section{Conclusions}\label{conclusion}
  We have studied the exponential decay of $N$ bosons with strong delta interactions from the double-well structure. Using the resonance expansion approach, we have analysed the effect of the splitting barrier on the decay rate of the $N$-particle system in dependence on $N$. We have found that    the decay rate of an initial state with  an odd number of particles is  strongly affected by the splitting barrier, in contrast to  its coherence which is insensitive to changes in the height of the splitting barrier.
Our results have shown that in the exponential regime the "radiating state" effectively reproduces the spatial correlations in the internal region.

% Uncomment the following two lines if you want to have a bibliography
%\bibliographystyle{alpha}
%\bibliography{document}

\end{document}